\documentclass[prl,twocolumn,showpacs,amssymb,superscriptaddress]{revtex4}

\begin{document}

\title{Microscopic Nuclear Structure Based upon a Chiral $NN$ Potential}
\author{L. Coraggio}
\affiliation{Dipartimento di Scienze Fisiche, Universit\`a 
di Napoli Federico II, \\ and Istituto Nazionale di Fisica Nucleare, \\
Complesso Universitario di Monte  S. Angelo, Via Cintia - I-80126 Napoli, 
Italy}
\author{A. Covello}
\affiliation{Dipartimento di Scienze Fisiche, Universit\`a 
di Napoli Federico II, \\ and Istituto Nazionale di Fisica Nucleare, \\
Complesso Universitario di Monte  S. Angelo, Via Cintia - I-80126 Napoli, 
Italy}
\author{A. Gargano}
\affiliation{Dipartimento di Scienze Fisiche, Universit\`a 
di Napoli Federico II, \\ and Istituto Nazionale di Fisica Nucleare, \\
Complesso Universitario di Monte  S. Angelo, Via Cintia - I-80126 Napoli, 
Italy}
\author{N. Itaco}
\affiliation{Dipartimento di Scienze Fisiche, Universit\`a 
di Napoli Federico II, \\ and Istituto Nazionale di Fisica Nucleare, \\
Complesso Universitario di Monte  S. Angelo, Via Cintia - I-80126 Napoli, 
Italy}
\author{T. T. S. Kuo}
\affiliation{Department of Physics, SUNY, Stony Brook, New York 11794}
\author{D. R. Entem}
\affiliation{Department of Physics, 
University of Idaho, Moscow, Idaho 83844}
\affiliation{Grupo de Fisica Nuclear, Universidad de Salamanca, 
37008 Salamanca, Spain}
\author{R. Machleidt}
\affiliation{Department of Physics, 
University of Idaho, Moscow, Idaho 83844}

\date{\today}

\begin{abstract}
We report on shell-model calculations employing
effective interactions derived from a new realistic nucleon-nucleon ($NN$)
potential based on chiral effective field theory.
We present results for $^{18}$O, $^{134}$Te, and $^{210}$Po. 
Our results are in excellent agreement with experiment
indicating a remarkable predictive power of the chiral $NN$ potential
for low-energy microscopic nuclear structure.
\end{abstract}


\maketitle

One of the most fundamental challenges pervading theoretical nuclear
physics for half a century is to understand the properties of
nuclei in terms of the basic interactions between 
the constituents.
After early progess~\cite{KB66}, the field was plagued for decades
by what is known as the off-shell uncertainty of the nuclear
force. Related to this issue is the problem that it was
not possible to derive the nuclear force from first
principles.

Recently, the picture changed dramatically when
the effective field theory (EFT) concept
was recognized in nuclear physics~\cite{EFT}.
The fundamental theory of strong interaction, QCD, is
nonperturbative in the low-energy regime characteristic for
nuclear physics; and this fact was generally perceived
as the great obstacle for a proper derivation of the
nuclear force. EFT shows the way out of this dilemma. 
The key is to notice that different phenomena in nature
are often characterized by different energy scales.
Traditional nuclear physics typically deals with 
low energies below the so-called chiral symmetry breaking scale,
$\Lambda_\chi \approx 1$ GeV, where the appropriate degrees of freedom 
are pions and
nucleons (and not quarks and gluons)
interacting via a force that is governed by
the symmetries of QCD, particularly, (broken) chiral symmetry.

The derivation of the nuclear force from chiral EFT
was initiated by Weinberg \cite{Wei90} and pioneered
by Ord\'o\~nez \cite{OK92} and van Kolck \cite{ORK94,Kol99}.
Subsequently, many groups got involved in the subject
\cite{RR94,KBW97,KGW98,Kai99,KSW98,EGM98}.
As a result, efficient methods for deriving
the nuclear force from chiral Lagrangians emerged
\cite{KBW97,KGW98,Kai99}
and the quantitative nature of the chiral $NN$ potential
improved \cite{EGM98}.
Nevertheless, for a long time, even the `best' chiral $NN$ potentials
were too inaccurate to serve as
reliable input for exact few-nucleon calculations or miscroscopic nuclear
many-body theory.
Recently, the situation has changed substantially with the 
appearance of the chiral $NN$ potential of Ref.~\cite{EM01},
also known as the Idaho chiral potential.
This potential reproduces the $NN$ data below 210 MeV with a $\chi^2$/datum
= 0.98~\cite{EMW01}, i.~e., with the same accuracy as
the high-precision $NN$ potentials constructed in the 
1990's \cite{Sto94,WSS95,MSS96,Mac01}.

The EFT approach inspires a new method \cite{bogner01} to renormalize the 
bare $NN$ interaction. The idea is to derive a low-momentum 
$NN$ potential, $V_{low-k}$, 
that preserves the physics of the original $NN$ interaction up to a
certain cut-off momentum $\Lambda$.
The deuteron binding energy, low-energy scattering phase shifts, and 
low-momentum half-on shell T-matrix of the original $V_{NN}$ are 
reproduced by $V_{low-k}$ \cite{bogner01}.
This is achieved by integrating out high-momentum components of
the original $V_{NN}$ by means of an iterative method 
\cite{andreozzi96,suzuki80}. 
Such decimation is similar to a Renormalization Group (RG) 
transformation \cite{WK74}.
The resulting $V_{low-k}$ is a smooth potential, which is suitable for 
being used in low-energy nuclear physics. 

We have employed the chiral $NN$ potential to conduct shell-model calculations 
for various two valence-particles nuclei.
More precisely, once we have derived the $V_{low-k}$, starting from
the chiral Idaho-B $NN$ potential~\cite{EM01}, we have employed it to 
calculate shell-model effective interactions using the $Q$-box plus folded
diagram method \cite{kuo80}.
These are the first microscopic nuclear structure calculations, for a
wide mass range, performed by using a new realistic $NN$ potential based 
on chiral effective field theory.

In order to illustrate how shell-model calculations based upon these
new chiral effective interactions can describe the spectroscopic properties 
of nuclei near closed shells, we report here results we have obtained 
for $^{18}$O, $^{134}$Te, and $^{210}$Po, which are specimens of
light-, medium-, and heavy-mass nuclei with two valence particles. 

As customary, we use single-particle energies extracted from the experimental 
spectra of the corresponding single-particle valence nuclei.
In Figs. 1-3 we compare the experimental \cite{nndc} and theoretical
spectra for $^{18}$O, $^{134}$Te, and $^{210}$Po, respectively.
More precisely, we consider the positive parity energy spectrum up to
4 MeV for $^{18}$O, while for $^{134}$Te and $^{210}$Po we report the
whole experimental spectra up to 5 and 3.3 MeV, respectively.

From Figs. 1-3 we see that the experimental spectra are very well
reproduced by the calculated ones, the discrepancy in the excitation
energies being less than 100 keV for most of the states.
As a matter of fact the rms deviation $\sigma$ \cite{sigma} turn out
to be 140, 111, and 86 keV for $^{18}$O, $^{134}$Te, and $^{210}$Po,
respectively. 

In Table I, we show the  observed \cite{audi93,fogelberg99} and calculated 
ground-state binding energies relative to the closest doubly closed core 
for the three nuclei under consideration.
For the absolute scaling of the sets of single-particle energies, the
mass excess values for nuclei with one particle with respect to
$^{16}$O, $^{132}$Sn, and $^{208}$Pb have been taken from
Ref. \cite{audi93,fogelberg99}.

For $^{134}$Te and $^{210}$Po, we assume that the contribution
of the Coulomb interaction between the valence 
protons is equal to the
matrix element of the Coulomb force between the states
$(g_{\frac{7}{2}})^2_{J^{\pi}=0^+}$ and
$(h_{\frac{9}{2}})^2_{J^{\pi}=0^+}$, respectively. From Table I we see
that our predictions are in very good agreement with experiment.

In summary, we have performed shell-model calculations in which
$V_{low-k}$ vertices derived from a chiral $NN$ potential (Idaho B) are
used as input instead of $G$ matrix vertices. 
The calculated spectra as well as the binding energies for the three
nuclei $^{18}$O, $^{134}$Te, and $^{210}$Po are in excellent agreement 
with the experimental data.
We wish to point out that the degree of accuracy is comparable to that
obtained in our previous studies using effective interactions
derived from modern realistic $NN$ potentials rooted in the meson
theory of nuclear forces, in particular the CD-Bonn potential \cite{inpc2001}.
We may conclude that our present calculations, which are the first
where a realistic chiral $NN$ potential has been used, show that this
potential is a valid input for a microscopic description of nuclear
structure properties.

\begin{acknowledgments}
This work was supported in part by the Italian Ministero dell'Universit\`a
e della Ricerca Scientifica e Tecnologica (MURST), by the European 
Social Fund (ESF), by the U.S. DOE Grant No.~DE-FG02-88ER40388,
by the U.S. NSF Grant No.~PHY-0099444, and by the Ram\'on Areces
Foundation (Spain). 
\end{acknowledgments}

\newpage

\begin{table}
\caption{Experimental and calculated ground-state binding energies
(MeV). See text for comments.}
\begin{ruledtabular}
\begin{tabular}{lccc}
 & & Binding energy & \\
 Nucleus & & & \\
 & Expt. &  & Calc. \\
\colrule
 & & & \\
 $^{18}$O & $12.19 \pm 0.00$ & & $12.19$ \\
 $^{134}$Te & $20.56 \pm 0.03$ & & $20.64$ \\
 $^{210}$Po & $~8.78 \pm 0.00$ & & $8.78$ \\
\end{tabular}
\end{ruledtabular}
\end{table}

\begin{figure}
\caption[ ]{Experimental and calculated spectrum of $^{18}$O.}
\caption[ ]{Experimental and calculated spectrum of $^{134}$Te.}
\caption[ ]{Experimental and calculated spectrum of $^{210}$Po.}
\end{figure}


\begin{references}
\bibitem{sigma} We define 
$\sigma =\{(1/N_{d}) \sum_{i}[E_{exp}(i)-E_{calc}(i)]^{2}\}^{1/2}$, where
$N_{d}$ is the number of data.
\end{references}
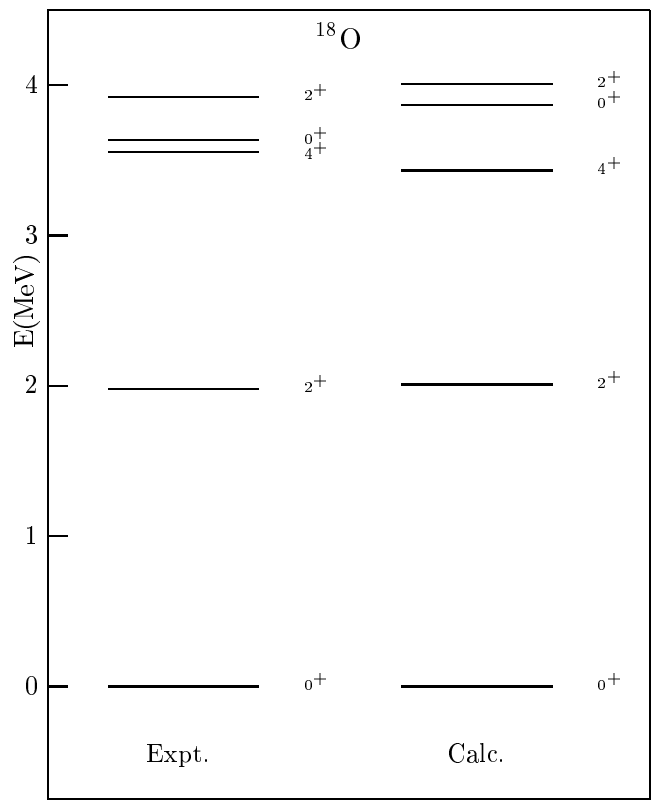
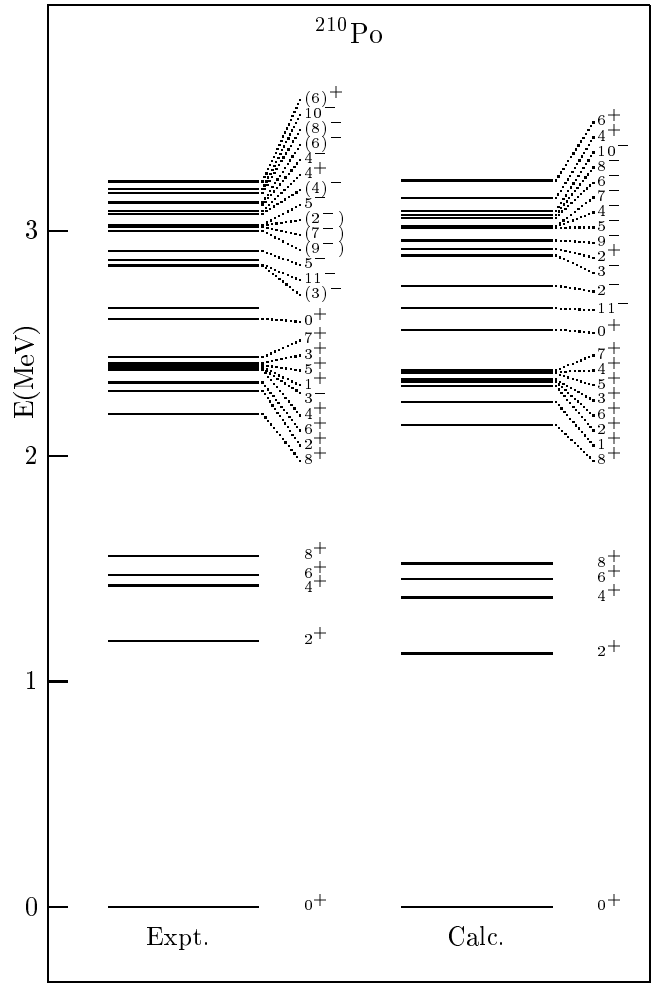
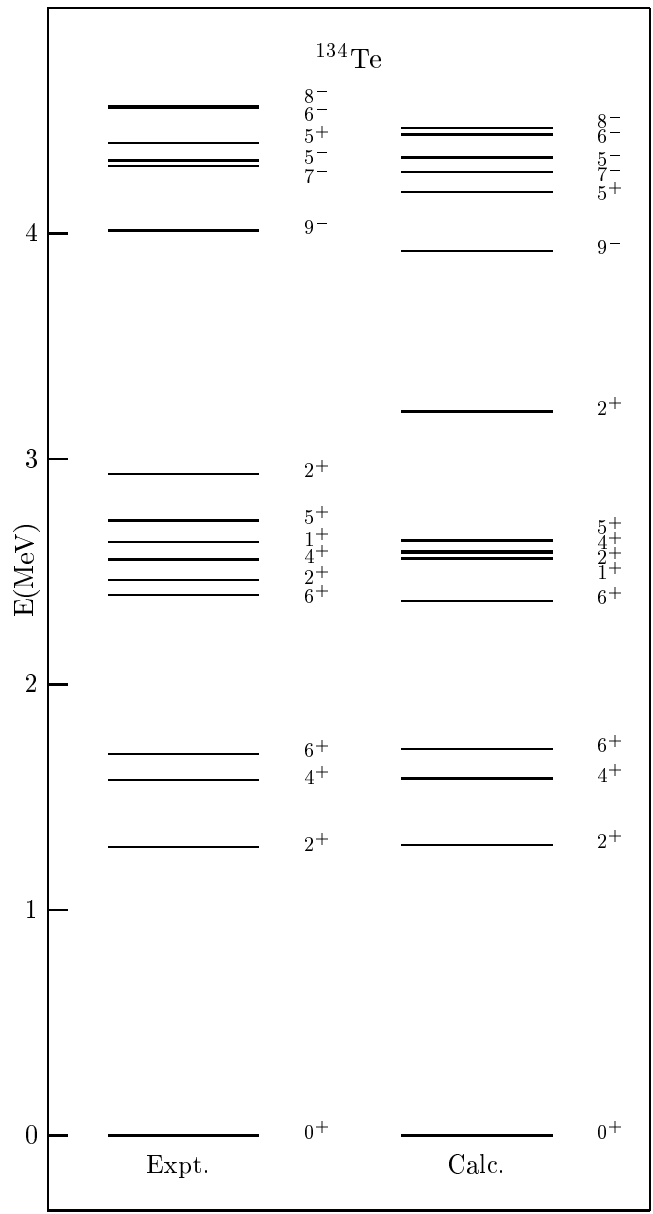
\end{document}